\documentstyle[12pt]{article}
\def\al{\alpha}
\def\be{\beta}
\def\ga{\gamma}
\def\de{\delta}
\def\ep{\epsilon}

\def\et{\eta}
\def\th{\theta}

\def\ka{\kappa}
\def\la{\lambda}

\def\si{\sigma}

\def\ta{\tau}

\def\ph{\phi}

\def\ch{\chi}
\def\ps{\psi}

\def\Ga{\Gamma}
\def\De{\Delta}
\def\Th{\Theta}

\def\mn{{\mu\nu}}
\def\cA{{\cal A}}
\def\cl{{\cal L}}
\def\fr#1#2{{{#1} \over {#2}}}
\def\prt{\partial}

\def\pt#1{\phantom{#1}}
\def\vev#1{\langle {#1}\rangle}
\def\bra#1{\langle{#1}|}
\def\ket#1{|{#1}\rangle}

\def\expect#1{\langle{#1}\rangle}

\def\half{{\textstyle{1\over 2}}}
\def\frac#1#2{{\textstyle{{#1}\over {#2}}}}

\def\lsim{\mathrel{\rlap{\lower4pt\hbox{\hskip1pt$\sim$}}
    \raise1pt\hbox{$<$}}}
\def\gsim{\mathrel{\rlap{\lower4pt\hbox{\hskip1pt$\sim$}}
    \raise1pt\hbox{$>$}}}
\def\sqr#1#2{{\vcenter{\vbox{\hrule height.#2pt
         \hbox{\vrule width.#2pt height#1pt \kern#1pt
         \vrule width.#2pt}
         \hrule height.#2pt}}}}

\def\lrprt{\stackrel{\leftrightarrow}{\partial}}
\def\lrprtnu{\stackrel{\leftrightarrow}{\partial^\nu}}

\newcommand{\beq}{\begin{equation}}
\newcommand{\eeq}{\end{equation}}
\newcommand{\bea}{\begin{eqnarray}}
\newcommand{\eea}{\end{eqnarray}}
\newcommand{\rf}[1]{(\ref{#1})}
 
\renewenvironment{thebibliography}[1]
 { \rm
   \begin{list}{\arabic{enumi}.}
    {\usecounter{enumi} \setlength{\parsep}{0pt}
     \setlength{\itemsep}{3pt} \settowidth{\labelwidth}{#1.}
     \sloppy
    }}{\end{list}}
 
\begin{document}
\titlepage
 
\begin{flushright}
{IUHET 354\\}
{December 1996\\}
\end{flushright}

\vglue 1cm
	    
\begin{center}
{{\bf CPT VIOLATION AND THE STANDARD MODEL 
\\}
\vglue 1.0cm
{Don Colladay and V. Alan Kosteleck\'y\\} 
\bigskip
{\it Physics Department, Indiana University\\}
\medskip
{\it Bloomington, IN 47405, U.S.A.\\}
 
\vglue 0.8cm
}
\vglue 0.3cm
 
\end{center}
 
{\rightskip=3pc\leftskip=3pc\noindent
Spontaneous CPT breaking arising in string theory
has been suggested as a possible 
observable experimental signature in neutral-meson systems.
We provide a theoretical framework
for the treatment of low-energy effects
of spontaneous CPT violation
and the attendant partial Lorentz breaking.
The analysis is within the context 
of conventional relativistic quantum mechanics 
and quantum field theory in four dimensions.
We use the framework to develop 
a CPT-violating extension to the minimal standard model
that could serve as a basis for
establishing quantitative CPT bounds.

}

\vskip 1 cm

\begin{center}
{\it Accepted for publication in Physical Review D\\}
{\it (scheduled for the June 1, 1997 issue) }
\end{center}

\vskip 1 cm

PACS: 11.30.Er, 12.60.-i, 11.25.-w

\newpage
 
\baselineskip=20pt
 
{\bf \noindent I. INTRODUCTION}
\vglue 0.4cm

Among the symmetries of the minimal standard model
is invariance under CPT.
Indeed,
CPT invariance holds under mild technical assumptions 
for any local relativistic point-particle field theory
\cite{s1}-\cite{s5}.
Numerous experiments have confirmed this result
\cite{pdg},
including in particular high-precision tests
using neutral-kaon interferometry 
\cite{c1,c2}.
The simultaneous existence of
a general theoretical proof of CPT invariance in particle physics 
and accurate experimental tests
makes CPT violation an attractive candidate signature
for non-particle physics such as string theory
\cite{kp1,kp2}.

The assumptions needed to prove the CPT theorem
are invalid for strings,
which are extended objects.
Moreover,
since the critical string dimensionality is larger than four, 
it is plausible that higher-dimensional
Lorentz breaking would be incorporated in a realistic model.
In fact,
a mechanism is known in string theory that can cause 
spontaneous CPT violation \cite{kp1}
with accompanying 
partial Lorentz-symmetry breaking \cite{ks}.
The effect can be traced to string interactions
that are absent in conventional four-dimensional
renormalizable gauge theory.
Under suitable circumstances,
these interactions can cause instabilities 
in Lorentz-tensor potentials,
thereby inducing spontaneous CPT and Lorentz breaking.
If in a realistic theory
the spontaneous CPT and partial Lorentz violation
extend to the four-dimensional spacetime,
detectable effects might occur 
in interferometric experiments with neutral kaons 
\cite{kp1,kp2},
neutral $B_d$ or $B_s$ mesons
\cite{kp2,ck1},
or neutral $D$ mesons 
\cite{kp2,ck2}.
For example,
the quantities parametrizing indirect CPT violation
in these systems could be nonzero.
There may also be implications for baryogenesis
\cite{bckp}.

In the present paper,
our goal is to develop 
within an effective-theory approach
a plausible CPT-violating extension of the minimal standard model
that provides a theoretical basis for establishing 
quantitative bounds on CPT invariance.
The idea is to incorporate notions of 
spontaneous CPT and Lorentz breaking
while maintaining the usual gauge structure
and properties like renormalizability.
To achieve this,
we first establish a conceptual framework and a procedure 
for treating spontaneous CPT and Lorentz violation
in the context of conventional quantum theory.
We seek a general methodology 
that is compatible with desirable features 
like microscopic causality
while being sufficiently detailed to permit explicit calculations. 

We suppose that underlying the effective four-dimensional action 
is a complete fundamental theory
that is based on conventional quantum physics 
\cite{fn1}
and is dynamically CPT and Poincar\'e invariant.
The fundamental theory is assumed to
undergo spontaneous CPT and Lorentz breaking.
In a Poincar\'e-observer frame
in the low-energy effective action,
this process is taken to fix the form
of any CPT- and Lorentz-violating terms.

Since interferometric tests of CPT violation are so sensitive,
we focus specifically on
CPT violation and the associated Lorentz-breaking issues
in a low-energy effective theory without gravity
\cite{fn2}.
For the most part,
effects from derivative couplings
and possible CPT-preserving but Lorentz-breaking terms 
in the action are disregarded,
and any CPT-violating terms
are taken to be small enough to avoid issues 
with standard experimental tests of Lorentz symmetry.
A partial justification for the latter assumption is that 
the absence of signals for CPT violation 
in the neutral-kaon system provides one of
the best bounds on Lorentz invariance
\cite{pdg}.

Our focus on the low-energy effective model 
bypasses various important theoretical issues 
regarding the structure of the underlying fundamental theory
and its behavior at scales above electroweak unification, 
including the origin and (renormalization-group) stability 
of the suppression of CPT breaking 
and the issue of mode fluctuations
around Lorentz-tensor expectation values.
Since these topics involve 
the Lorentz structure of the fundamental theory,
they are likely to be related to the difficult hierarchy problems 
associated with compactification and the cosmological constant.

The ideas underlying our theoretical framework 
are described in sect.\ II.
A simple model is used to illustrate 
concepts associated with CPT and Lorentz breaking,
including the possibility of eliminating 
some CPT-violating effects through field redefinitions.
The associated relativistic quantum mechanics
is discussed in sect.\ III.
Section IV contains a treatment of some issues 
in quantum field theory.
A CPT-violating extension 
of the minimal standard model is provided in sect.\ V,
and the physically observable subset of CPT-breaking terms 
is established.
We summarize in sect.\ VI.
Some of the more technical results are
presented in the appendices.

\vfill\eject
{\bf \noindent II. BASICS}
\vglue 0.4cm

{\bf \noindent A. Effective Model for Spontaneous CPT Violation}
\vglue 0.4cm

We begin our considerations with a simple model
within which many of the basic features of 
spontaneous CPT violation can be examined.
The model involves a single massive Dirac field $\ps (x)$
in four dimensions with lagrangian density
\beq
\cl = \cl_0 - \cl^{\prime}
\quad ,
\label{lagrangian}
\eeq
where $\cl_0$
is the usual free-field Dirac lagrangian for a fermion $\ps$
of mass $m$,
and where $\cl^{\prime}$ 
contains extra CPT-violating terms to be described below.
For the present discussion,
we follow an approach
in which the C, P, T and Lorentz properties
of $\ps$ are assumed to be conventionally
determined by the free-field theory $\cl_0$
and are used to establish the 
corresponding properties of $\cl^{\prime}$
\cite{sachs}.
This method is intrinsically perturbative,
which is particularly appropriate here 
since any CPT-violating effects must be small.
In subsection IIC,
we consider the possibility of alternative definitions
of C, P, T and Lorentz properties
that could encompass the full structure of $\cl$.

We are interested in possible forms of $\cl^\prime$
that could arise as effective contributions
from spontaneous CPT violation in a more complete theory.
To our knowledge,
string theory forms the only class of (gauge) theories 
in four or more dimensions that are quantum consistent,
dynamically Poincar\'e invariant,
and known to admit an explicit mechanism 
\cite{kp1}
for spontaneous CPT violation
triggered by interactions in the lagrangian.
However,
to keep the treatment as general as possible
we assume only that the spontaneous CPT violation
arises from nonzero expectation values 
acquired by one or more Lorentz tensors $T$,
so $\cl^\prime$ is taken to be an effective 
four-dimensional lagrangian
obtained from an underlying theory involving
Poincar\'e-invariant interactions of $\ps$ with $T$.
The discussion that follows is independent
of any specifics of string theory
and should therefore be relevant to a non-string model 
with spontaneous CPT violation,
if such a model is eventually formulated.

Even applying the stringent requirement 
of dynamical Poincar\'e invariance,
an unbroken realistic theory can in principle 
include terms with derivatives, powers of tensor fields,
and powers of various terms quadratic in fermion fields.
However,
any CPT-breaking term that is to be part of
a four-dimensional effective theory
must have mass dimension four.
In the effective lagrangian,
each combination of fields and derivatives
of dimension greater than four 
therefore must have a corresponding weighting factor 
of a negative power $-k$ of at least one mass scale $M$
that is large compared to the scale $m$ 
of the effective theory.
In a realistic theory with the string scenario,
$M$ might be the Planck mass 
or perhaps a smaller mass scale 
associated with compactification and unification. 
Moreover,
since the expectations $\vev{T}$ of the tensors
$T$ are assumed to be Lorentz and possibly CPT violating,
any terms that survive in $\cl^\prime$
after the spontaneous symmetry breaking
must on physical grounds be suppressed,
presumably by at least one power of $m/M$
relative to the scale of the effective theory.

A hierarchy of possible terms in 
$\cl^\prime$ thus emerges,
labeled by $k= 0, 1, 2, \ldots$.
Omitting Lorentz indices for simplicity,
the leading terms with $k\le 2$ have the schematic form 
\beq
\cl^\prime \supset \fr {\la}{M^k} \vev{T}
\cdot \overline{\ps} \Ga (i \prt )^k \ps + {\rm h.c.}
\quad .
\label{lcpt}
\eeq 
In this expression,
the parameter $\la $ is a dimensionless coupling constant,
$(i \partial)^k$ represents $k$ four-derivatives
acting in some combination on the fermion fields,
and $\Ga$ represents some gamma-matrix structure.
Terms with $k \ge 3$ 
and with more quadratic fermion factors also appear,
but these are further suppressed.
Note that contributions of the form \rf{lcpt}
arise in string theory
\cite{kp2}.
Note also that naive power counting 
indicates the dominant terms with $k\leq 1$ 
are renormalizable.

For $k=0$,
the above considerations indicate that
the dominant terms of the form \rf{lcpt}
must have expectations $\vev{T} \sim m^2/M$.
In the present work,
we focus primarily on this relatively simple case.
Most of the general features arising from CPT and Lorentz violation
together with some of our more specific results 
remain valid when terms with other values of $k$ are considered,
but it remains an open issue to investigate 
the detailed properties of
terms with $k=1$ and expectations $\vev{T} \sim m$
or those with $k=2$ and expectations $\vev{T} \sim M$.
Both these could in principle contribute 
leading effects in the low-energy effective action.

Each contribution to $\cl^\prime$
from an expression of the form \rf{lcpt}
is a fermion bilinear involving a $4\times 4$ spinor matrix $\Ga$.
Regardless of the complexity and number 
of the tensors $T$ inducing the breaking,
$\Ga$ can be decomposed as a linear combination
of the usual 16 basis elements of the gamma-matrix algebra.
Only the subset of these that produce CPT-violating bilinears
are of interest for our present purposes,
and they permit us 
to provide explicit and relatively simple expressions 
for the possible CPT-violating contributions to $\cl^\prime$.

For the case $k = 0$ of interest here,
we find two possible types of CPT-violating term:
\beq
\cl_a^\prime \equiv a_\mu \overline {\ps} \ga^{\mu} \ps 
\quad ,
\qquad
\cl_b^\prime \equiv b_\mu \overline {\ps} \ga_5 \ga^{\mu} \ps
\quad .
\label{abmu}
\eeq
For completeness,
we provide here also the terms appearing for 
the case $k = 1$, 
where we find three types of relevant contribution:
\beq
\cl_c^{\prime} \equiv \half i 
c^\al \overline{\ps} \lrprt_\al \ps 
\quad , \qquad 
\cl_d^{\prime} \equiv \half 
d^\al \overline{\ps} \ga_5 \lrprt_\al \ps 
\quad , \qquad 
\cl_e^\prime \equiv \half i 
e^\al_\mn \overline{\ps} \si^{\mn} \lrprt_\al \ps
\quad ,
\label{cde}
\eeq
where $A\lrprt_\mu B \equiv A \prt_\mu B - (\prt_\mu A) B$.
In all these expressions, 
the quantities 
$a_\mu$, $b_\mu$, $c^\al$, $d^\al$, and $e^\al_{\mn}$
must be real as consequences of their origins
in spontaneous symmetry breaking
and of the presumed hermiticity of the underlying theory.
They are combinations of coupling constants,
tensor expectations,
mass parameters,
and coefficients arising from the decomposition of $\Ga$.

In keeping with their interpretation 
as effective coupling constants 
arising from a scenario with spontaneous symmetry breaking,
$a_\mu$, $b_\mu$, $c^\al$, $d^\al$, and $e^\al_{\mn}$
are invariant under CPT transformations.
Together with the standard CPT-transformation properties 
ascribed to $\ps$,
this invariance causes the terms in
Eqs.\ \rf{abmu} and \rf{cde} to break CPT
\cite{fn3}.
As discussed above,
in the remainder of this work
we restrict ourselves largely to the expressions
in Eq.\ \rf{abmu}.

Allowing both kinds of term in Eq.\ \rf{abmu}
to appear in $\cl^\prime$
produces a model lagrangian of the form
\beq
\cl = \half i \overline{\ps} \ga^\mu \lrprt_\mu \ps 
- a_\mu \overline{\ps} \ga^\mu \ps
- b_\mu \overline{\ps} \ga_5 \ga^\mu \ps
- m \overline{\ps} \ps
\quad .
\label{modellag}
\eeq
The variational procedure generates a modified Dirac equation:
\beq
\left(i \ga^{\mu} \partial_{\mu} - a_{\mu} \ga^{\mu}
- b_{\mu} \ga_5 \ga^{\mu} - m \right) \ps = 0
\quad .
\label{mdeq}
\eeq

Associated with this Dirac-type equation
is a modified Klein-Gordon equation.
Proceeding with the usual squaring procedure, 
in which the Dirac-equation operator with opposite mass sign 
is applied to the Dirac equation from the left,
leads to the Klein-Gordon-type expression 
\beq
\left[ (i\prt - a)^2 - b^2 - m^2 
+ 2 i \ga_5 \si^{\mn} b_\mu (i \prt_\nu - a_\nu) \right] 
\ps (x) = 0
\quad .
\label{kgeq}
\eeq
This equation is second order in derivatives,
but unlike the usual Klein-Gordon case
it contains off-diagonal terms in the spinor space.
These may be eliminated by repeating the squaring procedure,
this time applying the operator in \rf{kgeq} 
with opposite sign for the off-diagonal piece.
The result is a fourth-order equation
satisfied by each spinor component
of any solution to the modified Dirac equation:
\beq
\left\{\left[ (i \prt - a)^2 - b^2 - m^2 \right]^2
+ 4 b^2 (i \prt - a)^2
- 4 \left[ b^\mu (i \prt_\mu - a_\mu) \right]^2 \right\} 
\ps (x) = 0
\label{scalareq}
\quad .
\eeq 

\vglue 0.6cm
{\bf \noindent B. Continuous Symmetries}
\vglue 0.4cm

Consider next the continuous symmetries
of the model with lagrangian \rf{modellag}.
For definiteness,
we begin with an analysis in a given oriented inertial frame
in which values of the quantities
$a_\mu$ and $b_\mu$ are assumed to have been specified.
The effects of rotations and boosts are considered later.

The CPT-violating terms in \rf{modellag}
leave unaffected the usual global U(1) gauge invariance,
which has conserved current 
$j^\mu = \overline \ps \ga^\mu\ps$.
Charge is therefore conserved in the model.
These terms also leave unaffected the usual
breaking of the chiral U(1) current 
$j_5^\mu = \overline \ps \ga_5\ga^\mu\ps$
due to the mass term.
In what follows,
we denote the volume integrals of the
current densities $j^\mu$ and $j_5^\mu$
by $J^\mu$ and $J_5^\mu$, respectively.

The model is also invariant under translations
provided the tensor expectations are assumed constant,
i.e.,  
provided the possibility of CPT-breaking soliton-type solutions 
in the underlying theory is disregarded.
This leads to a conserved canonical energy-momentum tensor
$\Th^{\mu\nu}$ given by
\beq
\Th^{\mu\nu} = \half i \overline{\ps} \ga^\mu 
\lrprtnu \ps 
\quad , \qquad 
\partial_{\mu} \Th^{\mu\nu} = 0 
\quad ,
\label{enmomtens}
\eeq
and a corresponding conserved four-momentum $P^\mu$.
These expressions have the same form as in the free theory.
Note,
however,
that constancy of the energy and momentum 
does not necessarily imply conventional behavior
under boosts or rotations.
Note also that the presence of the CPT-violating terms 
in the Dirac equation 
destroys the usual symmetrizability property of $\Th^{\mn}$.
The antisymmetric part $\Th^{[\mn ]}$ is
\beq
\Th^{[\mn ]}\equiv 
\Th^{\mn} - \Th^{\nu \mu} = 
- \frac 1 4 \partial_{\al} 
\left[\overline{\ps} \{ \ga^{\al} , \si^{\mn} \} \ps \right]
- a^{[\mu} j^{\nu ]} - b^{[\mu} j_{5}^{\nu ]}
\quad ,
\eeq 
which is no longer a total divergence.
The conventional construction of a symmetric energy-momentum tensor,
involving a subtraction of this antisymmetric part 
from the canonical energy-momentum tensor,
would affect the conserved energy and momentum
and is therefore presumably inapplicable in the present case.
The implications of this 
for a more complete low-energy effective theory 
that includes gravity remain to be explored.

Next, consider the effect of Lorentz transformations,
i.e., rotations and boosts.
Conventional Lorentz transformations in special relativity
relate observations made in two inertial frames with differing
orientations and velocities.
These transformations can be implemented as coordinate changes,
and we call them observer Lorentz transformations.
It is also possible to consider transformations
that relate the properties of two particles 
with differing spin orientation or momentum 
within a specific oriented inertial frame.
We call these particle Lorentz transformations.
For free particles under usual circumstances,
the two kinds of transformation are (inversely) related.
However,
this equivalence fails for particles under the action
of a background field.

The reader is warned to avoid confusing
observer Lorentz transformations
(which involve coordinate changes)
or particle Lorentz transformations
(which involve boosts on particles or localized fields
but \it not \rm on background fields)
with a third type of Lorentz transformation
that within a specified inertial frame  
boosts all particles and fields simultaneously,
including background ones.
The latter are sometimes called 
(inverse) active Lorentz transformations.
For the case of free particles,
they coincide with particle Lorentz transformations.
We have chosen to avoid applying the terms active and passive
here because they are insufficient to distinguish
the three kinds of transformation 
and because in any case 
their interpretation varies in the literature.

The distinction between observer and particle transformations
is relevant for the present model,
where the CPT-violating terms can be regarded
as arising from constant background fields 
$a_\mu$ and $b_\mu$.
The point is that these eight quantities
transform as two four-vectors
under observer Lorentz transformations
and as eight scalars under particle Lorentz transformations,
whereas they are coupled to currents that 
transform as four-vectors under both types of transformation.
This means that observer Lorentz symmetry is still
an invariance of the model,
but the particle Lorentz group is (partly) broken. 

Physical situations with features like this 
can readily be identified.
For example,
an electron with momentum perpendicular 
to a uniform background magnetic field moves in a circle.
Suppose in the same observer frame
we instantaneously increase the magnitude of the electron momentum
without changing its direction,
causing the electron to move in a circle of larger radius.
This (instantaneous) particle boost 
leaves the background field unaffected.
However,
if instead an observer boost 
perpendicular to the magnetic field is applied,
the electron no longer moves in a circle.
This is viewed in the new inertial frame
as an $E\times B$ drift caused by
the presence of an electric field.
In this example,
the background magnetic field transforms 
into a different electromagnetic field under observer boosts
but (by definition) is unchanged by particle boosts,
in analogy to the transformation of $a_\mu$ and $b_\mu$
in the CPT-violating model.

From the viewpoint of this example,
the unconventional aspect of the CPT-violating model 
is merely that the constant fields $a_\mu$ and $b_\mu$
are a global feature of the model.
They cannot be regarded as 
arising from localized experimental conditions,
which would cause them to transform 
under particle Lorentz transformations
as four-vectors rather than as scalars. 
The behavior of $a_\mu$ and $b_\mu$ as background fields
and hence as scalars under particle Lorentz transformations 
is a consequence of their origin as nonzero expectation values
of Lorentz tensors in the underlying theory.
These Lorentz-tensor expectations break 
those parts of the particle Lorentz group
that cannot be implemented as unitary transformations
on the vacuum.
This is in parallel with other situations involving
spontaneous symmetry breaking,
such as ones commonly encountered in the treatment
of internal symmetries.

The preservation of observer Lorentz symmetry
is an important feature of the model.
It is a consequence of observer Lorentz invariance
of the underlying fundamental theory.
This symmetry is unaffected 
by the appearance of tensor expectation values
by virtue of its implementation via coordinate transformations.
As an illustration of its use in the effective model, 
we show that it permits a further classification 
of types of CPT-violating term
according to the observer Lorentz properties of $a_\mu$ and $b_\mu$.
Thus,
for example,
if $b_\mu$ is future timelike in one inertial frame,
it must be future timelike in all frames.
This implies that a class of inertial frames can be found
in which $b_\mu = b (1,0,0,0)$,
where calculations are potentially simplified. 
A similar argument for the lightlike or spacelike cases
shows that the CPT-violating physics of the
four components of $b_\mu$ can in each case 
be reduced to knowledge of its Lorentz type
and a single number specifying its magnitude.
Inertial frames within this ideal class 
are determined by the little group of $b_\mu$,
which can in turn be used to simplify 
(partially) the form of $a_\mu$.

The reader is cautioned that the class of inertial frames
selected in this way may be distinct from 
experimentally relevant inertial frames such as,
for example,
those defined using the microwave background radiation 
and interpreting the dipole component
in terms of the motion of the Earth.
The point is that,
given an inertial frame,
the process of spontaneous Lorentz violation
in the underlying theory is assumed to produce
some values of $a_\mu$ and $b_\mu$.
In this specific inertial frame,
there is no reason \it a priori \rm 
why these values should take the ideal form
described above.
One is merely assured of the existence of
some frame in which the ideal form can be attained.

The current $J^{\la\mu\nu}$
for particle Lorentz transformations takes the usual form
when expressed in terms of the energy-momentum tensor:
\beq
J^{\la\mu\nu}=
x^{[\mu} \Th^{\la\nu ]}
+ \frac 1 4 \overline{\ps} \{ \ga^\la , \si^{\mn} \} \ps 
\quad .
\label{Lorcur}
\eeq
This current is conserved
at the level of the underlying theory
with spontaneous symmetry breaking,
but in the effective low-energy
theory where the spontaneous breaking 
appears as an explicit symmetry violation 
the conservation property is destroyed. 
In the latter case,
the corresponding Lorentz charges $M^{\mn}$ obey
\beq
\fr {dM^{\mn}}{dt} = 
- a^{[\mu} J^{\nu ]} - b^{[\mu} J_{5}^{\nu ]}
\quad .
\label{lviol}
\eeq

Given explicit values of $a_\mu$ and $b_\mu$
in some inertial frame,
Eq.\ \rf{lviol} can be used directly
to determine which Lorentz symmetries are violated.
Note that 
if either $a_\mu$ or $b_\mu$ vanishes,
the Lorentz group is broken to the little group
of the nonzero four-vector.
This means that the \it largest \rm
Lorentz-symmetry subgroup that can remain 
as an invariance of the model lagrangian \rf{modellag}
is SO(3), E(2), or SO(2,1).
Since $a_\mu$ and $b_\mu$ represent two four-vectors
in four-dimensional spacetime,
they define a two-dimensional plane.
Transformations involving the two orthogonal dimensions
have no effect on this plane.
This means that the \it smallest \rm
Lorentz-symmetry subgroup that can remain
is a compact or noncompact U(1).

In a realistic low-energy effective theory,
CPT-violating terms would break the particle Lorentz group
in a manner related to the breaking  
given by Eq.\ \rf{lviol}.
Since no zeroth-order CPT violation has been observed
in experiments,
CPT-violating effects in the string scenario 
are expected to be suppressed
by at least one power of the Planck mass
relative to the scale of the effective theory.
However,
the interesting and involved issue of exactly
how small the magnitudes of $a_\mu$ and $b_\mu$ 
(or their equivalents in a realistic model)
must be to satisfy current experimental constraints 
lies beyond the scope of the present work.
We confine our remarks here to noting 
that the partial breaking of particle Lorentz invariance
discussed above generates an effective boost dependence
in the CPT-breaking parameters.
This could provide a definite experimental signature 
for our framework
if CPT violation were detected at some future date.

\vglue 0.6cm
{\bf \noindent C. Field Redefinitions}
\vglue 0.4cm

For the discussions in the previous subsections,
we adopted a practical approach 
to the definition of CPT and Lorentz transformations.
It involves treating C, P, T and Lorentz properties of $\ps$
as being defined via the free-field theory ${\cal L}_0$
and subsequently using them to establish
the symmetry properties of ${\cal L}^{\prime}$.
This approach requires caution, 
however,
because in principle alternative definitions 
of the symmetry transformations could exist
that would leave the full theory $\cl$ invariant.

Consider first an apparently 
CPT- and Lorentz-violating model formed with $a_\mu$ only,
defined in a given inertial frame by the lagrangian
\beq
\cl [\ps ]= \cl_0 [\ps] - \cl_a^\prime [\ps ]
\quad .
\eeq
Introducing in this frame a field redefinition of $\ps$ 
by a spacetime-dependent phase,
\beq
\ch = \exp (i a \cdot x )\ps \quad ,
\eeq
the lagrangian 
expressed in terms of the new field is
$\cl [\ps 
= \exp (- i a \cdot x) \ch ]\equiv \cl_0 [\ch ]$.
This shows that the model 
is equivalent to a conventional free Dirac theory,
in which there is no CPT or Lorentz breaking,
and thereby provides an example of 
redefining symmetry transformations
to maintain invariance
\cite{fn4}.

The connection between the Poincar\'e generators
in the two forms of the theory can be found explicitly 
by substituting $\ps = \ps [\ch ]$ 
in the Poincar\'e generators for $\cl [\ps ]$
and extracting the combinations needed to 
reproduce the usual Poincar\'e generators for $\cl_0 [\ch ]$.
We find that the charge and chiral currents 
$j^\mu$ and $j_5^\mu$ 
take the same functional forms in both theories
but that the form of the canonical energy-momentum tensor changes,
\beq
\Th^{\mu\nu} 
= \half i \overline{\ch} \ga^\mu\lrprtnu\ch 
+ a^\nu \overline{\ch} \ga^\mu \ch
\quad ,
\eeq
producing a corresponding change in 
the Lorentz current $J^{\la\mu\nu}$.
This means that
in the original theory $\cl [\ps ]$
we could introduce modified Poincar\'e currents 
$\tilde\Th^{\mn}$ and $\tilde J^{\la\mu\nu}$
that have corresponding conserved charges 
generating an unbroken Poincar\'e algebra.
These currents are given as functionals of $\ps$ by
\beq
\tilde\Th^{\mn}= \Th^{\mn} - a^\nu j^\mu
\quad ,\qquad
\tilde J^{\la\mu\nu}= J^{\la\mu\nu} - x^{[\mu} a^{\nu ]} j^\la
\quad .
\eeq

The existence of this connection between the two
theories depends critically
on the existence of the conserved current $j^\mu$.
In the model \rf{modellag}
with both $a_\mu$ and $b_\mu$ terms,
the component $\cl_a^\prime$ can be eliminated
by a field redefinition as before
but there is no similar transformation removing $\cl_b^\prime$
because conservation of the chiral current $j_5^\mu$ is violated
by the mass.
In the massless limit of this model
the chiral current is conserved,
and we can eliminate both $a_\mu$ and $b_\mu$
via the field redefinition
\beq
\ch = \exp (i a \cdot x - i b \cdot x \ga_5) \ps
\quad .
\eeq
For the situation with $m\ne 0$,
however,
this redefinition would introduce 
spacetime-dependent mass parameters.

The term $\cl_a^\prime$ in Eq.\ \rf{abmu}
is reminiscent of a local U(1) coupling,
although there is no local U(1) invariance
in the theory \rf{modellag}.
It is natural and relevant to our later considerations
of the standard model
to ask how the above discussion of field redefinitions
is affected if the U(1) invariance of the 
original theory is gauged.
Then,
the term $\cl_a^\prime$ 
has the same form as a coupling to a constant background 
electromagnetic potential.
At the classical level,
this would be expected to have no effect
since it is pure gauge.
However,
a conventional quantum-field gauge transformation
involving both $\ps$ and the electromagnetic potential
$A_\mu$ cannot eliminate $a_\mu$,
since the theory is invariant under such transformations.
Instead,
the electromagnetic field can be taken as
the sum of a classical c-number background field $\cA_\mu$
and a quantum field $A_\mu$,
whereupon $a_\mu$ can be regarded as contributing to
an effective $\cA_\mu$.
Conventional classical gauge transformations
can be performed on the c-number potential $\cA_\mu$,
while leaving the quantum fields $\ps$ and $A_\mu$ unaffected.
This changes the lagrangian but should not change the physics.
In fact,
the resulting gauge-transformed lagrangian 
is unitarily equivalent to the original one
under a field redefinition on $\ps$ of the form
discussed above for the ungauged model.

To summarize,
in the gauged theory
the CPT-breaking term $\cl_a^\prime$
can be interpreted as a background gauge choice
and eliminated via a field redefinition 
as in the ungauged case.
We note in passing that
related issues arise for certain nonlinear gauge choices
\cite{dirac}
and in the context of efforts to 
interpret the photon as a Nambu-Goldstone boson
arising from (unphysical) spontaneous Lorentz breaking
\cite{heisenberg}-\cite{eguchi}.
In typical models of the latter type,
a four-vector bilinear condensate 
$\vev{\overline\ps\ga_\mu\ps}$
plays a role having some similarities to that of $a_\mu$.

The model \rf{modellag} involves only a single fermion field.
All CPT-violating effects can also be removed
from certain theories describing more than one fermion field
in which each fermion has a term of the form $\cl_a^\prime$.
For example,
this is possible if there is no fermion mixing 
and each such CPT-violating term 
involves the same value of $a_{\mu}$,
or if the fermions have no interactions or mixings that acquire
spacetime-dependence upon performing the field redefinitions.
However,
in generic multi-fermion theories with
CPT violation involving fermion-bilinear terms,
it is impossible to eliminate all CPT-breaking effects
through field redefinitions.
Nonetheless,
since lagrangian terms that spontaneously break CPT 
necessarily involve paired fermion fields,
at least one of the quantities $a_\mu$ can be removed.
This means that only differences between values of $a_\mu$
are observable.
Examples appear in the context of the CPT-violating extension
of the standard model discussed in section V.

\vglue 0.6cm
{\bf\noindent III. RELATIVISTIC QUANTUM MECHANICS}
\vglue 0.4cm

In this section,
we discuss some aspects of relativistic quantum mechanics
based on Eq.\ \rf{mdeq},
with $\ps$ regarded as a four-component wave function.
The results obtained provide further insight 
into the nature of the CPT-violating terms
and are precursors to the quantum field theory.
The analogous treatment in the context of the standard model 
involves several fermion fields,
for which CPT-violating terms of the form $\cl_a^\prime$
cannot be altogether eliminated.
We therefore explicitly include
the quantity $a_\mu$ in the following analysis,
even though it could be eliminated by a field redefinition
for the simple one-fermion case.
In fact,
the reinterpretation of negative-energy solutions
causes the explicit effects of $a_\mu$ to be more involved than
might otherwise be expected.

The modified Dirac equation \rf{mdeq} can be solved 
by assuming the usual plane-wave dependence,
\beq
\ps(x) = e^{-i\la_\mu x^\mu} w(\vec\la)
\label{planewave}
\quad .
\eeq 
In this equation,
$w(\vec\la)$ is a four-component spinor satisfying
\beq
\left(\la_\mu \ga^\mu - a_\mu \ga^\mu - b_\mu \ga_5 \ga^\mu
- m \right) w(\vec\la) = 0 
\quad .
\eeq
For a nontrivial solution to exist, 
the determinant of the matrix acting on $w(\vec \la)$
in this equation must vanish.  
This means that
$\la^\mu \equiv (\la^0, \vec \la)$,
where 
\cite{fn6}
$\la^0 = \la^0(\vec{\la})$, 
must satisfy the requirement 
\bea
\left[ (\la - a)^2 - b^2 - m^2 \right]^2
+ 4 b^2 (\la - a)^2
- 4 \left[ b^\mu (\la_\mu - a_\mu) \right]^2 
= 0
\quad .
\label{disp}
\eea
This condition can also be
obtained directly from Eq.\ \rf{scalareq}
and the assumption \rf{planewave}.

The dispersion relation \rf{disp}
is a quartic equation for 
$\la^0(\vec{\la})$.
Although the Euler reducing cubic has
a relatively elegant form,
in part because Eq.\ \rf{disp}
contains no term cubic in $\la^0$, 
the algebraic solutions to this equation
are not particularly transparent.
Even without examining the analytical results,
however,
certain features of the solutions can be established.
One is that all four roots must be real,
due to hermiticity of the quantum-mechanical hamiltonian
\beq
H\ps \equiv i \fr {\prt \ps} {\prt t} = \left( -i \ga^0 \vec\ga
\cdot \vec\nabla + a_\mu \ga^0 \ga^\mu
- b_\mu \ga_5 \ga^0 \ga^{\mu} + m \ga^0 \right) \ps
\quad .
\label{ham}
\eeq
Another stems from  
the invariance of the quartic under the interchange
$(\la_\mu - a_\mu) \to -(\la_\mu - a_\mu)$,
which implies that to each solution $\la^0_+(\vec\la)$
there corresponds a second solution $\la^0_-(\vec\la)$ 
given by 
\beq
\la^0_-(\vec\la)= 
- \la^0_+(- \vec\la + 2 \vec a ) + 2 a^0
\quad .
\label{dispsymm}
\eeq
This equation and the invariance of the quartic
under the interchange $b_\mu \to - b_\mu$
show that,
unlike the conventional Dirac case,
the magnitudes of the eigenenergies 
of the four roots all differ generically 
as a direct consequence of the CPT-violating terms
\cite{fn7}.

Another qualitatively different feature of the present model
is that under certain conditions
the roots $\la^0 (\vec{\la})$ 
of the dispersion relation can display cusps.
For a conventional dispersion relation,
the energy is a smooth function of each three-momentum component
for both timelike and spacelike four-momenta,
while there is a cusp at the origin for the lightlike case. 
By examining discontinuities in the
derivatives of the roots $\la^0 (\vec{\la})$ 
with respect to the components of $\vec \la$,
we have demonstrated that 
the criterion for cusps to appear 
in the present model with $m^2 > 0$
is that $b_\mu$ be timelike.
The derivation is most straightforward 
using observer Lorentz invariance
to select one of the canonical frames listed in Appendix B, 
for which exact solutions to the dispersion relations 
can be found.
The presence of cusps appears 
to have no directly observable consequences,
in part because their size is governed by the magnitude of $b_\mu$,
which is highly suppressed in a realistic situation. 

The assumption that the CPT-violating quantities 
$a_{\mu}$ and $b_{\mu}$ are small relative 
to the scale $m$ of the low-energy theory
implies the dispersion relation \rf{disp}
must have two positive-valued roots 
$\la^0_{+(\al)}(\vec{\la})$
and two negative-valued roots 
$\la^0_{- (\al)}(\vec{\la})$,
where $\al = 1,2$.
Since these roots are eigenvalues of the time-translation operator,
the corresponding wave functions
can be termed positive- and negative-energy states,
respectively.
Useful approximate solutions for
$\la^0_{\pm(\al)}(\vec{\la})$
that are valid to second order 
for arbitrary small $a_\mu$ and $b_\mu$
are given in Eq.\ \rf{disp2} of Appendix A.
Some exact solutions valid for various
important special cases are provided in Appendix B.

Within conventional relativistic quantum mechanics,
negative-energy states are deemed to be filled,
forming the Dirac sea.
When a negative-energy state is excited to a positive-energy one,
it leaves a hole appearing to be a particle with
opposite energy, momentum, spin, and charge
to that of the negative-energy state.
In the present model,
however, 
when a negative-energy state 
moving in a CPT-violating background 
with parameters $a_\mu$ and $b_\mu$
is excited to a positive-energy one,
it leaves a hole appearing to be a particle with
opposite values as before but 
moving in a CPT-violating background 
with parameters $-a_\mu$ and $b_\mu$ instead.
This is because the term $\cl_a^\prime$ 
is odd under charge conjugation.
The same effect can be seen explicitly by constructing
the charge-conjugate Dirac equation for the model.
We find
\beq
\left(i \ga^{\mu} \partial_{\mu} + a_{\mu} \ga^{\mu}
- b_{\mu} \ga_5 \ga^{\mu} - m \right) \ps^c = 0
\quad ,
\label{ccdeq}
\eeq
where as usual $\ps^c \equiv C {\overline\ps}^T$ 
and $C$ is the charge-conjugation matrix.

The eigenfunctions corresponding to the two negative eigenvalues 
$\la^0_{- (\al)}(\vec{\la})$
can be reinterpreted as positive-energy, 
reversed-momentum wave functions 
in the usual way.
We introduce momentum-space spinors 
$u^{(\al)} (\vec p )$, $v^{(\al)} (\vec p )$
via the definitions
\beq
\ps^{(\al)}(x) = \exp (-i p_u^{(\al)} \cdot x ) u^{(\al)} (\vec{p}) 
\quad , \qquad
\ps^{(\al)}(x) = \exp (+i p_v^{(\al)} \cdot x ) v^{(\al)} (\vec{p})
\quad ,
\label{momspin}
\eeq
where the four-momenta are given by 
\bea
p_u^{(\al)} & \equiv & (E_u^{(\al)}, \vec{p})
\quad , \qquad 
E_u^{(\al)}(\vec{p}) = \la^0_{+ (\al)}(\vec{p})
\quad ,
\nonumber \\
p_v^{(\al)} & \equiv & (E_v^{(\al)}, \vec{p})
\quad , \qquad 
E_v^{(\al)}(\vec{p}) = -\la^0_{- (\al)}(-\vec{p})
\quad .
\label{reinterp}
\eea
The general forms of 
$u^{(\al)} (\vec p )$ and $v^{(\al)} (\vec p )$
are given in Appendix A.
The relation between the spinors $u$ and $v$ is
determined by the charge-conjugation matrix and 
the charge-conjugate Dirac equation \rf{ccdeq}.
For example,
$u^{(2)} (\vec p, a_\mu, b_\mu )
\propto v^{(1) c} (\vec p, -a_\mu, b_\mu )$,
where the dependence on $a_\mu$ and $b_\mu$
has been explicitly restored for clarity. 
The symmetry \rf{dispsymm}
of the dispersion relation then connects
the two sets of energies by  
\beq
E_v^{(2,1)}(\vec{p}) 
= E_u^{(1,2)} (\vec{p} + 2 \vec{a}) - 2 a^0
\quad .
\label{ensymm}
\eeq
The exact eigenenergies for the various canonical cases
are provided in Appendix B,
while Appendix C contains explicit solutions
for the eigenspinors in the special case $\vec b = 0$.

The four spinors 
$u^{(\al)} (\vec p )$, $v^{(\al)} (\vec p )$
are orthogonal.
Their normalization can be freely chosen,
although imposing the condition $(\ps^c)^c = \ps$ 
provides a partial constraint.
Our choice leads to orthonormality conditions
given by
\bea 
u^{(\al) \dagger} (\vec{p}) u^{(\al^{\prime})}(\vec{p}) =
\de^{\al \al^{\prime}} \fr {E_u^{(\al)}} {m} 
\quad & , & \quad
v^{(\al) \dagger} (\vec{p}) v^{(\al^{\prime})}(\vec{p}) =
\de^{\al \al^{\prime}} \fr {E_v^{(\al)}} {m} 
\quad ,
\nonumber \\
u^{(\al) \dagger} (\vec{p}) v^{(\al^{\prime})}(-\vec{p}) = 0
\quad & , & \quad
v^{(\al) \dagger} (-\vec{p}) u^{(\al^{\prime})}(\vec{p}) = 0
\quad .
\label{orthonorm}
\eea
Note, however,
that the Lorentz breaking precludes a 
simple generalization of the orthonormality relations 
involving the Dirac-conjugate spinors 
$\overline{u}^{(\al)}(\vec{p})$
and $\overline{v}^{(\al)}(\vec{p})$
instead of the hermitian-conjugate spinors
${u}^{(\al) \dagger}(\vec{p})$
and ${v}^{(\al) \dagger}(\vec{p})$.
Equation \rf{orthonorm} produces the completeness relation
\beq
\sum_{\al = 1}^{2} \left[ 
\fr m {E_u^{(\al)}(\vec p)} u^{(\al)} (\vec{p}) 
\otimes u^{(\al) \dagger} (\vec{p}) 
+ \fr m {E_v^{(\al)}(-\vec p)} v^{(\al)} (-\vec{p}) 
\otimes v^{(\al) \dagger} (-\vec{p}) 
\right] = I 
\quad .
\label{comp}
\eeq 
We remark in passing that another useful result 
is the modified Gordon identity
\bea
\overline{u}^{(\al^\prime)}(\vec{p}^{~\prime}) \ga^\mu 
u^{(\al)} (\vec{p}) & = &
\fr 1 {2m} \overline{u}^{\al^\prime} (\vec{p}^{~\prime}) 
\left[ p_u^{~\prime (\al^{\prime}) \mu}
+ p_u^{(\al) \mu} - 2 a^\mu 
\right.
\nonumber \\
&& \quad\qquad\qquad
\left.
+ i \si^{\mn} \left( 
p_{u \nu}^{~\prime (\al^{\prime})} 
- p_{u \nu}^{(\al)} - 2 \ga_5 b_\nu 
\right) \right] u^{(\al)} (\vec{p})
\quad .
\label{Gordon}
\eea 

The general solution to the modified Dirac equation
\rf{mdeq} can be written as a superposition 
of the four spinors $u^{(\al)}$, $v^{(\al)}$:
\beq
\ps (x) = \int \fr {d^3 p} {(2 \pi )^3} 
\sum_{\al = 1}^{2} \left[
\fr m {E_u^{(\al)}} b_{(\al)} (\vec{p})
e^{-i p_u^{(\al)} \cdot x} u^{(\al)} (\vec{p}) +
\fr m {E_v^{(\al)}} d^*_{(\al)} (\vec{p})
e^{i p_v^{(\al)} \cdot x} v^{(\al)} (\vec{p}) \right]
\quad ,
\label{soln}
\eeq
where 
$b_{(\al)} (\vec{p})$, $d^*_{(\al)} (\vec{p})$
are the usual complex weights for the momentum expansion.
We remind the reader that in this expression
the $a_\mu$ and $b_\mu$ dependence of the energies 
and the spinors is understood. 

In the above expressions,
the four-momenta are eigenvalues of the translation operators
and hence are conserved quantities.
They therefore represent canonical energy and momentum
rather than kinetic energy and momentum.
A distinction of this type occurs in many physical systems,
such as a charged particle moving in an electromagnetic field.
In the present case this means,
for example,
that the canonical four-momenta are \it not \rm 
related to velocity as are the usual kinetic four-momenta
in special relativity.  
The actual relationship can be explored 
by using the velocity operator,
given in relativistic quantum mechanics by
$\vec{v} \equiv {d \vec{x}}/{dt} = i[H,\vec{x}] = \ga^0 \vec{\ga}$,
where $H$ is the hamiltonian \rf{ham}.
The expectation value of this operator for a given wave packet 
is the vector-current integral and
gives the (group) velocity of the packet.

As an explicit example,
consider the special case $\vec b = 0$,
for which the eigenenergies and eigenspinors 
are provided in Appendices B and C,
respectively. 
Suppose a wave packet of energy $E$ and momentum $\vec p$ 
is constructed as a superposition 
of positive-energy spin-up solutions.
A short calculation produces 
\beq
\expect{\vec{v}} = 
\expect{ \fr {(|\vec{p} - \vec{a}| - b^0)} {(E - a^0)}
         \fr {(\vec{p} - \vec{a})} {|\vec{p} - \vec{a}|} }
\quad .
\label{velexpect}
\eeq
It follows that the velocity is related 
to the energy and momentum by
\beq
\ga m = E - a^0 
\quad , \qquad
\ga m \vec v = 
\vec p - \vec a 
- b^0 \fr {(\vec{p} - \vec{a})} {|\vec{p} - \vec{a}|} 
\quad ,
\label{evrel}
\eeq
where $\ga = 1/{\sqrt{1 - v^2}}$ as usual.
These are just the usual special-relativistic results
shifted by (small) amounts controlled by the CPT-violating terms.
Note that four-momentum conservation shows that 
the wave-packet velocity is constant, as usual.

Even in conventional Dirac quantum mechanics,
the above notion of velocity involves subtleties
associated with the presence of negative-energy solutions. 
For example,
the velocity operator does \it not \rm commute 
with the usual Dirac hamiltonian.
In the present CPT-violating model,
additional subtleties arise.
For example,
it follows from the properties of the roots of the
dispersion relation \rf{disp} or from the above discussion
that for the special case of timelike $b_\mu$
the velocity near the origin
is not in one-to-one correspondence with the 
conserved momentum.
Perhaps of more interest is that in the general case 
the velocity operator in the energy basis 
has additional off-diagonal components,
even in the positive-energy sector.
For the example above,
these oscillate transverse to $(\vec p - \vec a)$
with relatively large period of order $b_0^{-1}$.
They provide time-independent corrections
to the velocity eigenvalues,
but only at order $b_0^2$.
The implications of these features for possible 
bounds on $b_\mu$ are therefore likely to be limited
but remain to be explored.

A related approach to the notion of velocity
is to take the derivative of the energy with respect 
to the momentum.
For the special case $\vec b = 0$,
this definition produces the same result as above
and moreover can be obtained without the explicit wave function.
It therefore provides a relatively simple method
of investigating velocity-related issues.
For example,
causality of the model is related to the restriction of 
the group velocity to below the velocity of light.
If causality is satisfied,
the criterion
\beq 
| v^j | \equiv \left| \fr {\prt E}{\prt p^j} \right|  < 1
\quad 
\label{caus}
\eeq
should be obeyed for each $j = 1,2,3$.
Observer Lorentz invariance makes it sufficient
to examine the various canonical cases.
Calculating with the expressions in Appendix B,
we find that the criterion \rf{caus}
is satisfied for all values of $a_\mu$ and $b_\mu$.
This supports the notion that causality is maintained.
Although the Lorentz breaking does affect quantum wave propagation,
it apparently is mild enough to avoid superluminal signals.

Our treatment in this section 
of the relativistic quantum mechanics
of a single fermion in the presence of CPT-violating terms
could be further developed
to allow for interactions with conventional applied fields,
along the lines of the usual Dirac theory.
In detail,
this lies outside the scope of the present work.
We remark,
however,
that standard Green-function methods should be applicable.
In particular,
we can introduce a generalized Feynman propagator 
$S_F (x-x^{\prime} )$
satisfying
\beq
\left(i \ga^{\mu} \partial_{\mu} - a_{\mu} \ga^{\mu}
- b_{\mu} \ga_5 \ga^{\mu} - m \right) 
S_F(x - x^{\prime}) = \de^4 (x - x^{\prime})
\label{fpe}
\quad 
\eeq
and obeying the usual Feynman boundary conditions.
It has integral representation
\beq
S_F (x-x^{\prime} ) = \int_{C_F} \fr {d^4 p} {(2 \pi)^4}
e^{-i p \cdot (x - x^{\prime})} 
\fr {1} 
{p_\mu\ga^\mu - a_\mu\ga^\mu - b_\mu \ga_5 \ga^\mu - m} 
\quad ,
\label{fp}
\eeq
where $C_F$ is the direct analogue 
of the usual Feynman contour in $p_0$ space,
passing below the two negative-energy poles
and above the positive-energy ones.
Appendix D contains some remarks about this propagator,
including a closed-form integration for the case $\vec b = 0$.

\vglue 0.6cm
{\bf\noindent IV. QUANTUM FIELD THEORY}
\vglue 0.4cm

In this section,
we discuss a few aspects of the quantum field theory
associated with the model lagrangian \rf{modellag}.
As in the usual Dirac case, 
direct canonical quantization is unsatisfactory,
and the quantization condition is found instead by imposing 
positivity of the conserved energy.

Promoting the Fourier coefficients in the expansion \rf{soln} 
to operators on a Hilbert space,
we can obtain from Eq.\ \rf{enmomtens}
an expression for the normal-ordered
conserved energy $P_0 = \int d^3x : \Th^0_{\pt{0} 0}:$.
This expression is positive definite for $a^0 < m$,
provided the following nonvanishing 
anticommutation relations are imposed:
\bea
\{b_{(\al)} (\vec{p}), b^{\dagger}_{(\al^{\prime})}
(\vec{p}^{~\prime}) \} & = & (2 \pi)^3 
\fr {E_u^{(\al)}} {m} 
\de_{\al \al^{\prime}}
\de^3 (\vec{p} - \vec{p}^{~\prime})   
\quad ,
\nonumber \\
\{d_{(\al)} (\vec{p}), d^{\dagger}_{(\al^{\prime})}
(\vec{p}^{~\prime}) \} & = & (2 \pi)^3 
\fr {E_v^{(\al)}} {m} 
\de_{\al \al^{\prime}}
\de^3 (\vec{p} - \vec{p}^{~\prime})   
\quad .
\eea
For simplicity,
the dependence on $a_\mu$ and $b_\mu$ is suppressed
in these and subsequent equations.
The corresponding equal-time field anticommutators are given by
\bea
\{ \ps_{j} (t,\vec{x}), \ps^{\dagger}_{k} 
(t, \vec{x}^{\prime}) \} &=& \de_{j k} \de^3 
(\vec{x} - \vec{x}^{\prime})
\quad ,
\nonumber \\
\{ \ps_{j} (t,\vec{x}), \ps_{k} 
(t, \vec{x}^{\prime}) \} & = & 
\{ \ps^{\dagger}_{j} (t,\vec{x}), \ps^{\dagger}_{k} 
(t, \vec{x}^{\prime}) \} = 0 
\quad ,
\label{etar}
\eea 
where the spinor indices $j$, $k$ are explicitly shown.

Using these expressions,
the normal-ordered conserved charge becomes 
\beq
Q = \int \fr {d^3 p} {(2 \pi)^3} 
\sum_{\al = 1}^2 \left[ 
\fr m {E_u^{(\al)}}
b^{\dagger}_{(\al)} (\vec{p}) b_{(\al)} (\vec{p}) -
\fr m {E_v^{(\al)}}
d^{\dagger}_{(\al)} (\vec{p}) d_{(\al)} (\vec{p}) 
\right]
\quad .
\eeq
Similarly,
the normal-ordered conserved four-momentum is
\beq
P_{\mu} = \int \fr {d^3 p} {(2 \pi)^3} 
\sum_{\al = 1}^2 \left[ 
\fr m {E_u^{(\al)}} p^{(\al)}_{u \mu}
b^{\dagger}_{(\al)} (\vec{p}) b_{(\al)} (\vec{p}) +
\fr m {E_v^{(\al)}} p^{(\al)}_{v \mu}
d^{\dagger}_{(\al)} (\vec{p}) d_{(\al)} (\vec{p}) 
\right]
\quad .
\eeq 
The reader can verify that
the operator $P_{\mu}$ generates spacetime translations
by determining the commutation relation with the field $\ps$:
$i[P_{\mu}, \ps(x)] = \partial_{\mu} \ps (x)$.

The creation and annihilation operators 
can be written in terms of the fields as:
\bea
b_{(\al)} (\vec{p}) & = & \int d^3 x~
e^{i p_u^{(\al)} \cdot x} 
\overline{u}^{(\al)} (\vec{p}) 
\ga^0 \ps (x) 
\quad , 
\nonumber \\
d^{\dagger}_{(\al)} (\vec{p}) & = & \int d^3 x~
e^{- i p_v^{(\al)} \cdot x} 
\overline{v}^{(\al)} (\vec{p}) 
\ga^0 \ps (x)
\quad .
\label{partops}
\eea
The vacuum state $\ket{0}$ of the Hilbert space obeys
\beq
b_{(\al)} (\vec{p}) \ket{0} = 0
\quad , \qquad
d_{(\al)} (\vec{p}) \ket{0} = 0
\quad .
\eeq
Acting on $\ket{0}$,
the creation operators 
$b^{\dagger}_{(\al)} (\vec{p})$ and
$d^{\dagger}_{(\al)} (\vec{p})$ 
produce particles and antiparticles with four-momenta
$p_u^{(\al) \mu}$ and $p_v^{(\al) \mu}$,
respectively.
The reinterpretation \rf{reinterp},
which is based on the usual heuristic arguments 
in relativistic quantum mechanics,
therefore makes sense in the field-theoretic framework.
As expected,
the usual fourfold degeneracy of the 
eigenstates of the hamiltonian for a given three-momentum 
is broken by the CPT-violating terms.

The above expressions can be used to establish 
various results for the field theory
with nonzero $a_\mu$ and $b_\mu$. 
For example,
the (time-dependent) commutation relation
between the conserved charges $P_{\mu}$ and 
the quantum operators 
$M^{\mn} = \int d^3x : J^{0\mu\nu} :$,
obtained from the operator form
of the currents \rf{Lorcur},
is found to be
\beq
i[P^{\la}, M^{\mn}] = 
- g^{\la [\mu} P^{\nu ]}
- g^{\la 0} (a^{[\mu} J^{\nu ]} + b^{[\mu} J_{5}^{\nu ]})
\quad ,
\eeq
where  
$J^{\mu} = \int d^3x : j^{\mu} :$ and
$J_5^{\mu} = \int d^3x : j_5^{\mu} :$
are integrals of the charge and chiral currents.
The $\la = 0$ component of this equation is the quantum-field
analogue of Eq.\ \rf{lviol}.

The generalizations of the
equal-time anticommutation relations \rf{etar}
to unequal times 
must be solutions to the modified Dirac equation 
in each variable and
must reduce to the usual results in the limit 
where $a_\mu$ and $b_\mu$ vanish.
The correct expressions can in principle be derived
by evolving forward in time one of the two fields in 
each anticommutator of Eq.\ \rf{etar}.
We write
\beq
\{ \ps (x), \overline {\ps} (x^\prime) \} 
= i S(x - x^\prime ) \quad ,\qquad
\{ \ps (x), {\ps} (x^\prime) \} = 
\{ \overline{\ps} (x), \overline {\ps} (x^\prime) \} = 0 
\quad ,
\label{utar}
\eeq 
where 
\beq
S (x-x^{\prime} ) = \int_{C} \fr {d^4 p} {(2 \pi)^4}
e^{-i p \cdot (x - x^{\prime})} 
\fr {1} 
{p_\mu\ga^\mu - a_\mu\ga^\mu - b_\mu \ga_5 \ga^\mu - m} 
\quad .
\label{af}
\eeq
In this expression,
$C$ is the analogue of the usual closed contour in $p_0$ space
encircling all the poles in the anticlockwise direction.
Some comments about this anticommutator function
are given in Appendix E,
along with a closed-form integration for $\vec b = 0$.
For this case,
we have checked explicitly that the anticommutators
\rf{utar} are determined by the integral \rf{af}.

The anticommutators \rf{utar}
are relevant to the causal structure of the quantum theory.
In particular,
for the case $\vec b = 0$
the results in Appendix E can be used to show
that the anticommutator of two fields separated 
by a spacelike interval vanishes:
\beq
\{ \ps_{\al} (x), \overline{\ps}_{\be} 
(x^{\prime}) \} = 0 \quad , \quad  
(x - x^{\prime})^2 < 0
\label{micro}
\quad .
\eeq
Observables constructed out of bilinear products of 
field operators and separated by a spacelike interval 
therefore commute.
This means that the quantum field theory 
with timelike $b_\mu$ preserves microscopic causality
in all associated observer frames.
The breaking of Lorentz invariance and the
distortions relative to conventional propagation
are apparently mild enough to exclude superluminal signals,
in agreement with the result from relativistic quantum mechanics.
The result might be anticipated
since observer Lorentz invariance holds 
and the particle Lorentz breaking involves 
only local terms in the lagrangian.
A direct analytical proof of microscopic causality 
in the quantum field theory 
for the cases of lightlike or spacelike $b_\mu$
would be of interest
but is hampered by the complexity of the integral \rf{af}.

We next turn to issues associated with interacting field theory.
For the most part,
since the CPT-violating terms in the model lagrangian 
can be treated exactly,
any added conventional interactions 
can be handled with standard methods.
In what follows,
we suppose that the Dirac fermion in the model 
has interactions with one or more other fields
that are of a type acceptable within a conventional approach,
and we discuss effects from CPT-violating terms.

Essentially all the standard assumptions 
underlying treatments of conventional interacting field theories
can reasonably be made in the present context.
Thus,
for example,
the property of observer Lorentz invariance 
ensures that consistent quantization
can be established in all observer frames
once it is established in a given frame.
Much of the usual analysis is performed in a given observer frame,
which in the present context 
means that the values of $a_\mu$ and $b_\mu$ are fixed.
Distinct effects are to be expected only in calculations 
for which particle Lorentz covariance plays 
an essential role.
For the most part,
matters proceed in a straightforward manner
at the level of the general framework of interacting field theory.
One exception we have found
is the explicit derivation of the K\"all\'en-Lehmann 
spectral representation for the vacuum expectation value of
the field anticommutator,
which normally takes advantage 
of both CPT and particle Lorentz covariance
\cite{kl1}-\cite{kl4}.
The spectral representation could be used to 
investigate microscopic causality of the interacting theory,
although the conventional local interactions we consider 
seem unlikely to introduce difficulties in this regard.
In any case,
it remains an open issue to obtain
the spectral representation in the present case,
where additional four-vectors appear in the theory.

The construction of the in and out fields 
and the definition of the $S$ matrix
can be implemented in the normal manner.
The LSZ reduction procedure for $S$-matrix elements 
generates expressions involving vacuum expectation values 
of time-ordered products of the interacting fields,
as usual,
but with external-leg factors for fermions 
involving the modified Dirac operator
$(i \ga^{\mu} \partial_{\mu} - a_{\mu} \ga^{\mu}
- b_{\mu} \ga_5 \ga^{\mu} - m )$
or its conjugate
$(i \ga^{\mu} {\stackrel{\leftarrow}\partial}_{\mu} 
+ a_{\mu} \ga^{\mu} + b_{\mu} \ga_5 \ga^{\mu} + m )$.

The canonical Dyson formalism for the perturbation series
of time-ordered products of interacting fields in terms of in fields
can be applied without encountering difficulties.
Standard expressions emerge,
including the vacuum bubbles.
Wick's theorem holds,
reducing the time-ordered products into normal-ordered products 
and pair contractions of in fields.
For fermion in fields, 
the vacuum expectation value of a pairwise contraction 
can be shown explicitly to be the
generalized Feynman propagator introduced in the previous section:
$\bra{0} T \ps (x) \overline \ps (x^{\prime}) \ket{0} 
= i S_F(x - x^{\prime})$. 
In a momentum-space Feynman diagram,
the corresponding propagator is 
\beq
S_F(p) = 
\fr 1 {p_\mu\ga^\mu - a_\mu\ga^\mu - b_\mu \ga_5 \ga^\mu - m} 
\quad .
\label{momprop}
\eeq
The momentum-space Feynman rules require 
this modified propagator for internal fermion lines 
and modified spinors on external fermion lines,
but are otherwise unchanged from those of a conventional theory.
For example,
translational invariance insures that energy and momentum
are conserved in any process
and so the standard four-momentum delta functions emerge.

Since the CPT-violating terms are renormalizable
by naive power counting,
we anticipate no difficulties 
with the usual renormalization program.
Details of loop calculations 
lie beyond the scope of the present work
and remain an interesting open issue.
We remark in passing 
that the form \rf{dirp} of the propagator given in Appendix D
shows that $a_{\mu}$ cancels around all closed fermion loops
in analogy with Furry's theorem.

We also expect the unitarity of the $S$ matrix to be unaffected
by CPT-violating terms.
Since a complete Hilbert-space solution exists
in the pure fermion case,
there are no hidden or inaccessible states
that could generate nonunitarity
of the type appearing
in the first stage of Gupta-Bleuler quantization
of quantum electrodynamics,
for example.
Moreover,
the interaction hamiltonian is hermitian.
In any event,
any nonunitarity 
appearing in a realistic model based
on a unitary fundamental theory would presumably be a signal 
that the domain of validity of the effective low-energy theory 
is being breached.

In determining physically relevant quantities
such as cross sections or transition rates,
kinematic factors appear.
For the most part,
these are straightforward to obtain.
A subtlety arises in the calculation 
of a physical cross section
because the standard definition involves 
the notion of incident flux
defined in terms of incoming particle velocity.
Since the velocity-momentum relation has corrections
involving $a_\mu$ and $b_\mu$
(cf.\ Eq.\ \rf{evrel}),
there are corresponding small modifications 
to the kinematic factors in standard cross-section formulae
expressed in terms of conserved momenta.
In a realistic case,
these are unobservable because they are suppressed.
This should be contrasted with the CPT-violating corrections
arising in an amplitude from the modified fermion propagator,
which are also suppressed 
but might be detected in interferometric experiments
using neutral-meson oscillations
\cite{kp2}.

\vglue 0.6cm
{\bf \noindent V. STANDARD-MODEL EXTENSION}
\vglue 0.4cm

In this section,
we consider the possibility of generalizing  
the minimal standard model by adding CPT-violating terms
within a self-consistent framework
of the type described in the previous sections.
Since CPT violation has not been observed in nature,
any CPT-violating constants appearing 
in an extension of the minimal standard model must be small.
In what follows,
we assume that these constants are
singlets under the unbroken gauge group,
but as before behave under particle Lorentz transformations
as tensors with an odd number of Lorentz indices.
Our primary goal is to obtain an explicit and realistic model
for CPT-violating interactions
that could serve as a basis for establishing  
quantitative CPT bounds.

The discussion in the previous sections 
is limited to CPT-breaking fermion bilinears.
However,
other types of terms violating CPT in the lagrangian could 
in principle originate from spontaneous symmetry breaking.
We adopt here a general approach,
investigating possible CPT-violating extensions 
to the minimal standard model
such that the SU(3) $\times$ SU(2) $\times$ U(1) gauge structure 
is maintained.
To preserve naive power-counting renormalizability
at the level of the unbroken gauge group,
we restrict attention to terms involving field operators
of mass dimension four or less.
The simultaneous requirements of 
gauge invariance, suitable mass dimensionality, and CPT violation
allow relatively few new terms in the action 
\cite{fn8}.

Any lagrangian term must be formed from combinations 
of covariant derivatives 
and fields for leptons, quarks, gauge bosons, and Higgs bosons.
We consider first allowed CPT-violating lagrangian extensions 
involving fermions.
Inspection shows that the only possibilities 
satisfying the above criteria are pure fermion-bilinear terms
without derivatives
\cite{fn9}.
In the present context involving many fermions,
the analysis given in the previous sections
of such terms requires some generalization.
Since SU(3) invariance precludes quark-lepton couplings,
we can treat the lepton and quark sectors separately
as usual.

Consider first the lepton sector.
Denote the left- and right-handed lepton multiplets by
\beq
L_A = \left( \begin{array}{c} \nu_A \\ l_A
\end{array} \right)_L
\quad , \quad
R_A = (l_A)_R
\quad ,
\eeq
where
\beq
\ps_R \equiv \frac 1 2 ( 1 + \ga_5 ) \ps
\quad , \qquad
\ps_L \equiv \frac 1 2 ( 1 - \ga_5 ) \ps
\quad ,
\label{handproj}
\eeq
as usual,
and where $A = 1,2,3$ labels the lepton flavor:
$l_A \equiv (e, \mu, \ta)$, 
$\nu_A \equiv (\nu_e, \nu_\mu, \nu_\ta)$.
Then,
the most general set of 
CPT-violating lepton bilinears 
consistent with gauge invariance is
\beq
\cl^{\rm CPT}_{\rm lepton} = 
- (a_L)_{\mu AB} \overline{L}_A \ga^{\mu} L_B
- (a_R)_{\mu AB} \overline{R}_A \ga^{\mu} R_B
\label{cptviol}
\quad .
\eeq
The constant flavor-space matrices $a_{L,R}$ are hermitian.
The presence of the $\ga^\mu$ factor
allows fields of only one handedness to appear
in a given term while maintaining gauge invariance.
This contrasts with conventional Yukawa couplings,
in which fields of both handedness appear 
and invariance is ensured by the presence of the Higgs doublet.

After spontaneous symmetry breaking, 
the mass eigenstates (denoted with carets) 
are constructed with standard unitary transformations
\beq
{\nu}_{LA} = (U^\nu_L)_{AB} \hat{\nu}_{LB}
\quad , \quad 
l_{LA} = (U^l_L)_{AB} \hat{l}_{LB} 
\quad , \quad 
l_{RA} = (U^l_R)_{AB} \hat{l}_{RB} 
\quad .
\eeq
The CPT-violating term in Eq.\ (\ref{cptviol}) becomes
\beq
\cl^{\rm CPT}_{\rm lepton} = 
- (\hat a_{{\nu}L})_{\mu AB} 
\overline{\hat{\nu}}_{LA} \ga^{\mu} \hat{\nu}_{LB}
- (\hat a_{{l}L})_{\mu AB} 
\overline{\hat{l}}_{LA} \ga^{\mu} \hat{l}_{LB} 
- (\hat a_{{l}R})_{\mu AB} 
\overline{\hat{l}}_{RA} \ga^{\mu} \hat{l}_{RB} 
\quad ,
\label{lepton1}
\eeq
where each matrix of constants $\hat a_\mu$ is obtained from
the corresponding $a_\mu$ 
via unitary rotation with the corresponding matrix $U$:
$\hat a_{\mu} = U^\dagger a_{\mu} U$.

Not all the couplings $\hat a_\mu$ are observable.
The freedom to redefine fields
allows some couplings to be eliminated,
in analogy to the discussion of section IIC 
for the model lagrangian.
Consider,
for example,
general field redefinitions of the form
\beq
\tilde{\nu}_{LA} = (V^\nu_L)_{AB} \hat{\nu}_{LB}
\quad , \quad 
\tilde l_{LA} = (V^l_L)_{AB} \hat{l}_{LB} 
\quad , \quad 
\tilde l_{RA} = (V^l_R)_{AB} \hat{l}_{RB} 
\quad ,
\label{redef}
\eeq
where the matrices $V(x^\mu)$ are unitary in generation space
and have the form $V=\exp(iH_\mu x^\mu)$
with $H_\mu$ hermitian.
Then,
in each kinetic term of the generic form 
$i\overline{\hat\ps} \ga^\mu\prt_\mu \hat\ps$,
the redefinition \rf{redef}
generates an apparent CPT-violating term of the form 
$-\overline{\tilde \ps}\ga^\mu 
\exp(iH_\la x^\la) H_\mu \exp(-iH_\nu x^\nu) \tilde\ps$.
A suitable choice of $H_\mu$
might therefore remove a CPT-violating term 
involving $\hat a_\mu$ from Eq.\ \rf{lepton1}.
The matrices $V$ must be chosen to leave unaffected 
the Yukawa couplings and 
the conventional nonderivative couplings
defining the neutrino fields as weak eigenstates.
It can be shown that 
$V^l_L$ must be a diagonal matrix of phases,
with $V^l_R =V^l_L$
and $V^\nu_L=(U^l_L)^\dagger U^\nu_L V^l_L$.
The freedom therefore exists to redefine the fields
so as to eliminate, say,
the three diagonal elements of the CPT-violating matrix 
in the neutrino sector. 
Note that the existence of this choice 
obviates possible theoretical issues 
arising from the combination of massless fields at zero temperature
and small negative energy shifts induced by CPT-breaking terms
\cite{fn10}.

Thus,
omitting tildes and carets,
the general CPT-violating extension 
of the lepton sector of the minimal standard model 
has the form
\beq
\cl^{\rm CPT}_{\rm lepton} = 
- (a_\nu)_{\mu AB} 
\overline{\nu}_A \half (1+ \ga_5) \ga^{\mu} \nu_B
- (a_l)_{\mu AB} \overline{l}_A \ga^{\mu} l_B
- (b_l)_{\mu AB} \overline{l}_A \ga_5 \ga^{\mu} l_B
\quad ,
\label{cptlepton}
\eeq
where we have used Eq.\ \rf{handproj}
to replace left- and right-handed couplings
with vector and axial vector couplings,
and where $(a_\nu)_{\mu AA} = 0$.
Note that Eq.\ \rf{cptlepton}
includes terms breaking individual lepton numbers,
although total lepton number remains conserved.
Flavor-changing transitions therefore exist in principle 
but are unobservable if
the CPT-violating couplings are sufficiently suppressed.
For example,
a fractional suppression of order $10^{-17}$ 
or smaller might occur in the string scenario
\cite{kp2}.

Consider next the quark sector.
The SU(3) symmetry 
ensures that all three quark colors of any given flavor
must have the same CPT-violating current coupling.
We can therefore disregard the color space in what follows,
and a construction analogous to that for the lepton sector
can be applied.
Denote the left- and right-handed components of the quark fields by
\beq
Q_A = \left( \begin{array}{c} u_A \\ d_A \end{array} \right)_L
\quad , \quad U_A = (u_A)_R \quad , \quad D_A = (d_A)_R
\quad , 
\eeq
where $A = 1,2,3$ labels the quark flavors:
$u_A \equiv (u,c,t)$, $d_A \equiv (d,s,b)$.
Then,
at the unbroken-symmetry level,
the most general CPT-violating coupling is 
\beq
\cl^{\rm CPT}_{\rm quark} = 
- (a_Q)_{\mu AB} \overline{Q}_A \ga^{\mu} Q_B
- (a_U)_{\mu AB} \overline{U}_{A} \ga^{\mu} U_{B} 
- (a_D)_{\mu AB} \overline{D}_A \ga^{\mu} D_B
\quad .
\label{cptviolq}
\eeq
As before, the constant flavor-space matrices $a_{Q,U,D}$
are hermitian.

After spontaneous symmetry breaking the mass eigenstates are
obtained with the standard unitary transformations
\bea
u_{LA} = (U^u_L)_{AB} \hat{u}_{LB} \quad & , & \quad
u_{RA} = (U^u_R)_{AB} \hat{u}_{RB} \quad , \quad
\nonumber \\
d_{LA} = (U^d_L)_{AB} \hat{d}_{LB} \quad & , & \quad 
d_{RA} = (U^d_R)_{AB} \hat{d}_{RB} \quad .
\eea
The CPT-violating expression \rf{cptviolq} becomes 
\bea
\cl^{\rm CPT}_{\rm quark} &=& 
-(\hat a_{uL})_{\mu AB} 
\overline{\hat{u}}_{LA} \ga^{\mu} \hat{u}_{LB} 
-(\hat a_{dL})_{\mu AB} 
\overline{\hat{d}}_{LA} \ga^{\mu} \hat{d}_{BL} 
\nonumber \\
&&
-(\hat a_{uR})_{\mu AB} 
\overline{\hat{u}}_{RA} \ga^{\mu} \hat{u}_{RB}
-(\hat a_{dR})_{\mu AB} 
\overline{\hat{d}}_{RA} \ga^{\mu} \hat{d}_{RB}
\quad .
\eea
As in the lepton sector,
each constant matrix $\hat a_\mu$ is obtained from
the corresponding $a_\mu$ via the appropriate unitary rotation.

Again,
field redefinitions can be used to 
eliminate some CPT violation.
Consider field redefinitions of the form
\bea
\tilde{u}_{LA} = (V^u_L)_{AB} \hat{u}_{LB}
\quad & , & \quad
\tilde u_{RA} = (V^u_R)_{AB} \hat{u}_{RB} 
\quad , \quad 
\nonumber \\
\tilde d_{LA} = (V^d_L)_{AB} \hat{d}_{LB} 
\quad & , & \quad
\tilde d_{RA} = (V^d_R)_{AB} \hat{d}_{RB} 
\quad ,
\label{qredef}
\eea
where as before the matrices $V=\exp(iH_\mu x^\mu)$
are unitary in generation space.
In this case,
invariance of the Yukawa and nonderivative couplings,
including the Cabbibo-Kobayashi-Maskawa mixings,
requires the effect of the matrices $V$ to reduce to 
multiplication by a single phase.
For example,
we can choose this phase so that the condition 
$(a_{uL})_{\mu 11}= (a_{uR})_{\mu 11} $ holds.
This removes the CPT-breaking vector coupling
from the u-quark sector.

Omitting tildes and carets,
the general CPT-violating extension 
of the quark sector of the minimal standard model is therefore
\beq
\cl^{\rm CPT}_{\rm quark} =
-(a_u)_{\mu AB} \overline{u}_A \ga^{\mu} u_B 
-(b_u)_{\mu AB} \overline{u}_A \ga_5 \ga^{\mu} u_B
-(a_d)_{\mu AB} \overline{d}_A \ga^{\mu} d_B 
-(b_d)_{\mu AB} \overline{d}_A \ga_5 \ga^{\mu} d_B ~~ ,
\label{cptquark}
\eeq
where 
$(a_u)_{\mu 11} \equiv 0$.
Again,
these terms include small flavor-changing effects
that are unobservable if the suppression is sufficiently small,
such as that of fractional order $10^{-17}$ or smaller
possible in the string scenario.
In contrast,
the diagonal contributions
might be detected in interferometric experiments 
that measure the phenomenological 
parameters $\de_P$ for indirect CPT violation
in oscillations of neutral-$P$ mesons,
where $P$ is one of $K$, $D$, $B_d$, or $B_s$.
Each quantity $\de_P$ is proportional to  
the difference between the diagonal elements
of the effective hamiltonian governing the 
time evolution of the corresponding $P$-$\overline P$ system.
Explicit expressions for $\de_P$ in terms of 
quantities closely related to those in Eq.\ \rf{cptquark}
have been given in Ref.\ \cite{kp2}.

The two equations \rf{cptlepton} and \rf{cptquark}
represent allowed CPT-violating extensions
of the fermion sector of the minimal standard model.
Next, 
we briefly consider 
other CPT-violating terms without fermions.

The only CPT-violating term involving the Higgs field
and satisfying our criteria 
is a derivative coupling of the form
\beq
\cl^{\rm CPT}_{\rm Higgs}
= i k^{\mu} \ph^{\dagger} D_{\mu} \ph + {\rm h.c.} 
\quad ,
\label{cpthiggs}
\eeq
where $k^\mu$ is a CPT-violating constant,
$D_\mu$ is the covariant derivative,
and $\ph$ is the usual SU(2)-doublet Higgs field.
Let us proceed under the assumption 
that no self-consistency issues arise 
for a scalar field that breaks CPT and Lorentz invariance,
so that standard methods apply.
Then,
Eq.\ \rf{cpthiggs} represents a contribution to
the Higgs-$Z^0_\mu$ sector of the model.
Disregarding possible CPT-preserving but Lorentz breaking
contributions to the static potential,
it can be shown that the term \rf{cpthiggs} 
produces a (stable) modification 
of the standard symmetry-breaking pattern 
to include an expectation value for the $Z^0_\mu$ field
with magnitude proportional to $k_\mu$.
Several kinds of effect ensue but if, 
as expected, 
the quantities $k^\mu$ are sufficiently small
then it can be shown that the results are either unobservable
or produce additional contributions 
to the fermion-bilinear terms already considered.

It is also possible to find CPT-violating terms
satisfying our criteria and involving only the gauge fields.
They are of the form
\bea
\cl^{\rm CPT}_{\rm gauge} &=&
  k_{3\ka} \ep^{\ka\la\mu\nu} 
{\rm Tr} (G_\la G_{\mu\nu} + \frac 2 3 G_\la G_\mu G_\nu)
+ k_{2\ka} \ep^{\ka\la\mu\nu} 
{\rm Tr} (W_\la W_{\mu\nu} + \frac 2 3 W_\la W_\mu W_\nu)
\nonumber\\ &&
+ k_{1\ka} \ep^{\ka\la\mu\nu} B_\la B_{\mu\nu} 
+ k_{0\ka} B^\ka 
\quad ,
\label{cptgauge}
\eea
where
$k_{3\ka}$, $k_{2\ka}$, $k_{1\ka}$, and $k_{0\ka}$ 
are CPT-violating constants.
Here,
$G_\mu$, $W_\mu$, $B_\mu$ are the (matrix-valued)
SU(3), SU(2), U(1) gauge bosons,
respectively,
and $G_{\mu\nu}$, $W_{\mu\nu}$, $B_{\mu\nu}$ 
are the corresponding field strengths.
The first three of these terms can be shown to leave unaffected 
the symmetry-breaking pattern,
and we expect only unobservable effects for  
small CPT-violating constants.
The field entering the term with coupling $k_{0\ka}$ 
is of dimension one.
It appears to produce a linear instability in the theory
because it involves the photon,
in which case it cannot emerge from
a fundamental theory with a stable ground state.

\vglue 0.6cm
{\bf\noindent VI. SUMMARY}
\vglue 0.4cm

In this paper,
we have developed a framework 
for treating spontaneous CPT and Lorentz breaking
in the context of conventional effective field theory.
The underlying action is assumed to be consistent 
and fully CPT and Poincar\'e invariant,
with solutions exhibiting spontaneous CPT and Lorentz breaking.
The effective low-energy field theory 
then remains translationally invariant 
and covariant under changes of observer inertial frame,
but violates CPT and partially breaks covariance 
under particle boosts.

Our focus has primarily been 
on lagrangian terms that involve
CPT-violating fermion bilinears,
which are relevant for experiments bounding CPT 
in meson interferometry.
In principle,
these terms can be treated exactly because they are quadratic.
We have investigated the relativistic quantum mechanics
and the quantum field theory 
of a model for a Dirac fermion involving CPT violation.
The analysis suggests that 
effective field theories with spontaneous CPT breaking 
have desirable properties 
like microscopic causality and renormalizability.
The existence of consistent theories of this type is reasonable
since they are analogous to conventional field theories
in a nonvanishing background.
Additional interactions appear minimally affected
by the CPT violation,
and the effects are largely restricted to modifications 
on fermion lines. 

Within the framework developed,
we have constructed a CPT-violating
generalization of the minimal standard model
that could be used in establishing quantitative CPT bounds.
The criteria of gauge invariance and power-counting renormalizability 
constrain the extension to a relatively simple form,
involving the extra terms given in 
Eqs.\ \rf{cptlepton}, \rf{cptquark},
\rf{cpthiggs}, and \rf{cptgauge}.
It has been previously been suggested 
\cite{kp1,kp2}
that the properties of neutral-meson systems $P\overline{P}$,
where $P$ is one of $K$, $D$, $B_d$, or $B_s$, 
are well suited to interferometric tests 
of spontaneous CPT violation,
with the experimentally measurable parameters
for (indirect) CPT violation 
being explicitly related to certain diagonal elements 
of the quark-sector CPT-breaking matrices 
given in Eq.\ \rf{cptquark}.
Investigating the current experimental constraints 
on the other CPT-violating parameters introduced here
is an interesting open topic
and could lead to additional signals for CPT violation.

\vglue 0.6cm
{\bf\noindent ACKNOWLEDGMENTS}
\vglue 0.4cm
We thank Robert Bluhm for discussion.
This work was supported in part
by the United States Department of Energy 
under grant number DE-FG02-91ER40661.

\vglue 0.6cm
{\bf\noindent APPENDIX A: EIGENSPINORS OF THE DIRAC EQUATION}
\vglue 0.4cm

Treating the CPT-violating parameters $a_\mu$ and $b_\mu$
as small relative to $m$,
the four roots $\la^0_{\pm(\al)}(\vec\la)$,
$\al = 1,2$,
of the dispersion relation \rf{disp} 
are given to second order by
\bea
\la^0_{\pm(\al)}(\vec\la)
& = & \pm \Biggr\{ m^2 + (\vec{\la} -
\vec{a})^2 
\nonumber \\ & & \qquad
+ 2 (-1)^\al \biggr[b_0^2 (\vec{\la} - \vec{a})^2
+ \vec{b}^2 m^2 + (\vec{b} \cdot (\vec{\la} - \vec{a}))^2 
\nonumber \\ & & 
\qquad \qquad \qquad \qquad \qquad
\mp 2 b_0 \vec{b} \cdot (\vec{\la} - \vec{a}) \sqrt{m^2 +
(\vec{\la} - \vec{a})^2} \biggr]^{\frac 1 2} 
\nonumber \\ & & \qquad
+ b_0^2 + \vec{b}^2 
\mp \fr {2 b_0 \vec{b} \cdot (\vec{\la} - \vec{a})}
{\sqrt{m^2 + (\vec{\la} - \vec{a})^2}} \Biggr\}^{\fr 1 2}
+ a_0
\quad .
\label{disp2}
\eea
This equation produces exact solutions to the dispersion relation
in any of the special cases for which
$b_0\vec b\cdot (\vec\la - \vec a) = 0$.
The eigenenergies of the four spinors
$u^{(\al)}(\vec{p})$, $v^{(\al)}(\vec{p})$
defined in Eq.\ \rf{momspin}
can be obtained by combining Eq.\ \rf{disp2}
with Eq.\ \rf{reinterp}.

In the general case,
the four spinor eigensolutions can be written 
in the Pauli-Dirac representation as 
\bea
u^{(\al)}(\vec{p}) & = & N_u^{(\al)} 
\left( \begin{array}{c}
\ph^{(\al)} \\
X_u^{(\al)} \ph^{(\al)}
\end{array} \right) 
\quad ,
\nonumber \\
v^{(\al)}(\vec{p}) & = & N_v^{(\al)} 
\left( \begin{array}{c}
X_v^{(\al)} \ch^{(\al)}
\\
\ch^{(\al)}
\end{array} \right) 
\quad .
\label{general}
\eea
In the first of these equations,
$N_u^{(\al)}$ 
is an arbitrary spinor normalization factor
and $X_u^{(\al)}$ is a spinor matrix defined by  
\beq
X_u^{(\al)} = 
\fr
{(E_u^{(\al)} - a_0 + m + \vec b \cdot \vec \si )
[(\vec p - \vec a ) \cdot \vec \si - b_0]}
{ (E_u^{(\al)} - a_0 + m)^2  - \vec{b}^2 }
\quad .
\label{xu}
\eeq
The analogous quantity $X_v^{(\al)}$
for the second equation
in \rf{general} can be found by 
replacing all subscripts $u$ by $v$ in Eq.\ \rf{xu}
and implementing the substitutions
$a_\mu \to -a_\mu$, 
$b_\mu \to -b_\mu$
wherever these quantities explicitly appear.
The quantities 
$\ph^{(\al)}$ and $\ch^{(\al)}$ are two-component spinors
satisfying the eigenvalue equations
\beq
\vec\ka_u^{(\al)} \cdot \vec\si \ph^{(\al)} = 
\et_u^{(\al)} \ph^{(\al)} 
\quad , \qquad
\vec\ka_v^{(\al)} \cdot \vec\si \ch^{(\al)} = 
\et_v^{(\al)} \ch^{(\al)} 
\quad ,
\eeq
with $[\vec\ka_u^{(\al)}]^2 = [\et_u^{(\al)}]^2$.
Here,
the vector $\vec\ka_u^{(\al)}$ 
and the scalar $\et_u^{(\al)}$ are given by
\bea
\vec\ka_u^{(\al)} &=& 
2 \left[ (p_{u\mu} - a_\mu) b^\mu + m b_0 \right]
(\vec p - \vec a) 
-\left [ (p_u - a)^2 + b^2 + m^2 
  + 2 m (E_u^{(\al)} - a_0) \right] \vec b 
\quad , \nonumber \\
\et_u^{(\al)} & = & 
2(E_u^{(\al)} - a_0) \vec b^2 
- 2 b_0 \vec b \cdot (\vec p - \vec a) 
- (E_u^{(\al)} - a_0 + m)
\left [ (p_u - a)^2 - b^2 - m^2 \right] 
\quad . \nonumber\\
\eea
The analogous quantities with subscripts $v$
are given by the same substitutions as before.

\vglue 0.6cm
{\bf\noindent APPENDIX B: EXACT EIGENENERGIES FOR CANONICAL CASES}
\vglue 0.4cm

For the case where $b_\mu$ is timelike,
observer Lorentz invariance can be used to select 
a canonical frame in which $\vec b = 0$.
In this frame,
we find the exact eigenenergies after reinterpretation are
\bea
E_u^{(\al)} & = & 
\left[ m^2 + (|\vec p - \vec a| + (-1)^\al b_0)^2 \right]^{1/2} 
+ a_0 
\quad ,
\nonumber \\
E_v^{(\al)} & = & 
\left[ m^2 + (|\vec p + \vec a| - (-1)^\al b_0)^2 \right]^{1/2} 
- a_0 
\quad ,
\label{b0en}
\eea
where $\al = 1,2$ as usual.

For the case of spacelike $b_\mu$, 
an observer frame can be chosen in which $b^0 = 0$.
After reinterpretation the exact eigenenergies become
\bea
E_u^{(\al)} & = & 
\left[ m^2 + (\vec p - \vec a)^2 
+ (-1)^\al 2 \sqrt{m^2 \vec b^2 
+ (\vec b \cdot (\vec p - \vec a))^2} 
+ \vec b^2 
\right]^{1/2} + a_0
\quad ,
\nonumber \\
E_v^{(\al)} & = & 
\left[ m^2 + (\vec p + \vec a)^2 
- (-1)^\al 2 \sqrt{m^2 \vec b^2 
+ (\vec b \cdot (\vec p + \vec a))^2} 
+ \vec b^2 
\right]^{1/2} - a_0
\quad .
\label{vecben}
\eea

Finally,
for the lightlike case $b_{\mu}b^{\mu} = 0$ the 
exact eigenvalues of the dispersion relation 
after reinterpretation are
\bea
E_u^{(\al)} & = & 
\left[ m^2 + (\vec p - \vec a - (-1)^\al \vec b)^2 \right]^{1/2} 
+ a_0 + (-1)^\al b_0 
\quad ,
\nonumber \\
E_v^{(\al)} & = & 
\left[ m^2 + (\vec p + \vec a + (-1)^\al \vec b)^2 \right]^{1/2} 
- a_0 - (-1)^\al b_0 
\quad .
\label{lighten}
\eea
These last expressions hold in all observer frames.

\vglue 0.6cm
{\bf\noindent APPENDIX C: EXPLICIT SOLUTION FOR $\vec b =0$}
\vglue 0.4cm

For the special case $\vec b = 0$,
the eigenenergies are given by Eq.\ \rf{b0en}
and the eigenspinors can be written in a relatively simple form.
Introducing momentum-space spinors via Eq.\ \rf{momspin},
we find in the Pauli-Dirac basis the expressions
\bea
u^{(\al)} (\vec{p}) & = & 
\sqrt{\fr {E_u^{(\al)}(E_u^{(\al)} - a_0 + m)} 
  {2 m (E_u^{(\al)} - a_0)}} 
\left( \begin{array}{c} 
\ph^{(\al)}(\vec p - \vec a) \\
\fr {- (-1)^\al | \vec{p} - \vec{a} | - b_0 } 
{E_u^{(\al)} - a_0 + m} 
  \ph^{(\al)}(\vec p - \vec a) 
\end{array} \right) 
\quad , \nonumber \\
v^{(\al)} (\vec{p}) & = & 
\sqrt{\fr {E_v^{(\al)}(E_v^{(\al)} + a_0 + m)} 
  {2 m (E_v^{(\al)} + a_0)}} 
\left( \begin{array}{c} 
\fr {- (-1)^\al | \vec{p} + \vec{a} | + b_0 } 
{E_v^{(\al)} + a_0 + m} 
  \ph^{(\al)}(\vec p + \vec a) \\
\ph^{(\al)}(\vec p + \vec a) 
\end{array} \right) 
\quad ,
\label{b0spin}
\eea
where $\al = 1,2$ as usual
and where we have chosen the normalization of the spinors
so that Eq.\ \rf{orthonorm} is satisfied.
In Eq.\ \rf{b0spin},
the two two-component spinors $\ph^{(\al)}(\vec \la)$
are the eigenvectors of $\vec\si \cdot \hat\la$
with eigenvalues $-(-1)^\al$.
If the spherical-polar angles
that $\vec\la$ subtends are specified as $(\th,\ph)$,
then the spinors $\ph^{(\al)}(\vec \la)$ are given explicitly by
\beq
\ph^{(1)}(\vec \la) = 
\left( \begin{array}{c} 
\cos{\fr \th 2} \\
\sin{\fr \th 2} e^{i \ph} 
\end{array} \right) \quad, \quad
\ph^{(2)}(\vec \la) = 
\left( \begin{array}{c}
- \sin{\fr \th 2} e^{-i \ph} \\
\cos{\fr \th 2} 
\end{array} \right)
\quad .
\eeq
Note that the structure of the CPT-violating terms
forces the spinors \rf{b0spin}
and their generalizations in Appendix A
to involve helicity-type states. 
In the limit of vanishing CPT violation,
the solutions \rf{b0spin} reduce to standard Dirac spinors
in the helicity basis.

\vglue 0.6cm
{\bf\noindent APPENDIX D: PROPAGATOR FUNCTIONS}
\vglue 0.4cm

It can be shown that the generalized Feynman propagator 
determined by Eqs.\ \rf{fpe} and \rf{fp} has the form 
\bea
S_F(x - x^{\prime}) &=& 
\left(i \ga^{\la} \prt_{\la} - a_{\la} \ga^{\la}
- b_{\la} \ga_5 \ga^{\la} + m \right) 
\left(i \ga^{\mu} \partial_{\mu} - a_{\mu} \ga^{\mu}
+ b_{\mu} \ga_5 \ga^{\mu} + m \right) 
\nonumber\\
&& \quad \qquad \qquad
\times \left(i \ga^{\nu} \partial_{\nu} - a_{\nu} \ga^{\nu}
+ b_{\nu} \ga_5 \ga^{\nu} - m \right) 
\De_F(x - x^{\prime}) 
\quad ,
\eea
where 
\beq
\De_F(y)
= \int_{C_F} \fr {d^4 p} {(2 \pi)^4}
e^{-i p \cdot y} 
\fr {1} 
{\left[ (p - a)^2 - b^2 - m^2 \right]^2
+ 4 b^2 (p - a)^2
- 4 \left[ b^\mu (p_\mu - a_\mu) \right]^2 }
\quad ,
\label{sfp}
\eeq
with $C_F$ the same contour in the $p_0$ plane as
that in Eq.\ \rf{fp}.

Direct integration for the special case $\vec b = 0$ gives
\bea
S_F(x - x^{\prime}) &=& 
e^{-i a \cdot (x - x^{\prime})}
\left(i \ga^{\mu} \partial_{\mu} 
+ b_{0} \ga_5 \ga^{0} + m \right) 
\nonumber\\
&& \qquad \qquad
\times \left( -\prt^2 - m^2 - b_0^2 
+ 2i b_0 \ga_5 \ga^{0} \ga^{j} \prt_j \right) 
\De_F(x - x^{\prime}) 
\quad ,
\label{dirp}
\eea
where
\beq
\De_F(x - x^{\prime}) = \fr 1 {16 \pi^2}
\fr {\sin{b_0 r}}{b_0 r}
\cases{ 
2i K_0(m \sqrt{r^2 - t^2})\quad , 
& $r^2 > t^2$\quad , \cr 
\pi H_0^{(2)} (m \sqrt{t^2 - r^2})\quad , 
& $r^2 < t^2$ \quad , \cr }
\eeq
where $r$ is the radial spherical-polar coordinate.
In this expression,
$K_0$ is a modified Bessel function and
$H_0^{(2)}$ is a Hankel function of the second kind.
The result \rf{dirp}
reduces to the standard one in the limit $a_\mu = b_0 = 0$.
Note that the propagator is singular on the light cone,
as usual. 

\vglue 0.6cm
{\bf\noindent APPENDIX E: ANTICOMMUTATOR FUNCTIONS}
\vglue 0.4cm

The anticommutator function $S(x - x^\prime)$
defined in Eq.\ \rf{af}
can be shown to be given by
\bea
S(x - x^{\prime}) &=& 
\left(i \ga^{\la} \partial_{\la} - a_{\la} \ga^{\la}
- b_{\la} \ga_5 \ga^{\la} + m \right) 
\left(i \ga^{\mu} \partial_{\mu} - a_{\mu} \ga^{\mu}
+ b_{\mu} \ga_5 \ga^{\mu} + m \right) 
\nonumber\\
&& \quad \qquad \qquad
\times \left(i \ga^{\nu} \partial_{\nu} - a_{\nu} \ga^{\nu}
+ b_{\nu} \ga_5 \ga^{\nu} - m \right) 
\De(x - x^{\prime}) 
\quad ,
\eea
where 
\beq
i \De(y)
= \int_{C} \fr {d^4 p} {(2 \pi)^4}
e^{-i p \cdot y} 
\fr {1} 
{\left[ (p - a)^2 - b^2 - m^2 \right]^2
+ 4 b^2 (p - a)^2
- 4 \left[ b^\mu (p_\mu - a_\mu) \right]^2 }
\quad ,
\label{saf}
\eeq
with $C$ being the contour of Eq.\ \rf{af}
in the $p_0$ plane.

For the special case $\vec b = 0$, 
direct integration gives
\bea
S(x - x^{\prime}) &=& 
e^{- i a \cdot (x - x^{\prime})}
\left(i \ga^{\mu} \partial_{\mu} 
+ b_{0} \ga_5 \ga^{0} + m \right) 
\nonumber\\
&& \quad \qquad \qquad
\times \left( -\prt^2 - m^2 - b_0^2 
+ 2i b_0 \ga_5 \ga^{0} \ga^{j} \prt_j \right) 
\De(x - x^{\prime}) 
\quad ,
\label{ae}
\eea
where
\beq
i \De(x - x^{\prime}) = 
- \fr 1 {8 \pi} \fr {\sin{b_0 r}}{b_0 r}
\cases{ 
J_0 (m \sqrt{t^2 - r^2}) \quad , & $t > r$ \quad , \cr 
0 \quad , & $-r<t<r$ \quad , \cr
-J_0 (m \sqrt{t^2 - r^2})\quad , & $t < - r$ \quad , \cr} 
\eeq
where $r$ is the radial spherical-polar coordinate
and $J_0$ is a Bessel function.
The expression \rf{ae} reduces to the standard result 
in the limit $a_\mu = b_0 = 0$.

\vglue 0.6cm
{\bf\noindent REFERENCES}
\vglue 0.4cm


\begin{thebibliography}{xx}
\bibitem{fn1}
A different source of CPT violation than that discussed here,
involving unconventional quantum mechanics,
has been suggested in the context of quantum gravity
\cite{qg1}-\cite{qg4}.
The experimental signature it might produce in 
the kaon system is known to be different
from that of the sources discussed in the present work 
\cite{qg5}.
\bibitem{fn3}
The effective coupling constants are also invariant
under the individual discrete transformations C, P, T.
The C, P, T properties of the various terms 
in Eqs.\ \rf{abmu} and \rf{cde} 
are therefore determined by
the standard C, P, T transformation properties of $\ps$.
\bibitem{fn4}
At the quantum level,
there is a unitarily implementable equivalence 
between the Hilbert spaces of the two theories.
It is generated by $U = \exp(i a\cdot x J^0)$.
This Hilbert-space map
acts to modify the phase of each state in the 
original CPT- and Lorentz-violating theory $\cl [\ps ]$
in a position- and charge-dependent way.
Neutral states, 
including the vacuum,
remain unaffected.
\bibitem{fn6}
The roots $\la^0(\vec{\la})$ 
also have arguments $a_\mu$ and $b_\mu$,
which for brevity are omitted throughout.
\bibitem{fn7}
These differences can intuitively be regarded as
particle-antiparticle and spin splittings.
Although details lie outside our present scope,
we note that features of this type 
suggest a variety of feasible experiments 
that could be used to bound 
CPT-violating parameters in an extension of the standard model
(cf.\ section V).
\bibitem{fn9}
If dimension-five operators are admitted
at the unbroken-symmetry level,
then among the additional terms appearing
after the SU(2) $\times$ U(1) breaking
could be corresponding dimension-five 
CPT-violating terms involving the Higgs field 
with derivative couplings to fermion-bilinear terms.
If such terms are deemed acceptable,
the lagrangian could then also contain 
dimension-four CPT-violating fermion-bilinear terms 
with (covariant) derivatives and couplings proportional to 
the dimensionless ratio of the Higgs expectation value
to a large mass scale.
Such derivative terms would be standard-model
generalizations of those appearing in Eq.\ \rf{cde}
and we expect their treatment to be relatively straightforward,
although details of the associated framework 
along lines discussed for the nonderivative case 
in previous sections would need to be established.
\bibitem{fn10}
Neutrino-mass terms in a nonminimal standard model
would exclude some of the redefinitions
but would also avoid any zero-mass issues.
\end{thebibliography}
\end{document}